\documentclass[aps,prl,twocolumn,groupedaddress]{revtex4-1}
\usepackage{amsmath,amssymb,graphicx}

\begin{document}

\title{Extraction of depth moments by exploiting the partial coherence of radiation}



\author{M. A. Beltran$^{1,2}$, T. C. Petersen$^{2}$, M. J. Kitchen$^{2}$, D. M. Paganin$^{2}$}

\affiliation{$^{1}$EMPA Swiss Federal Laboratories for Material Science and Technology, Switzerland.}

\affiliation{$^{2}$School of Physics and Astronomy, Monash University, Clayton, Victoria, 3800, Australia.}

\date{\today}

\begin{abstract}
We retrieve depth information (moments) of an object using partially coherent fields and defocus induced holographic contrast. Our analysis leads to a form of tomography that does not require sample or source rotation. The tomography method presented here is performed with only two in-line images.
\end{abstract}


\maketitle


In--line holographic imaging is one of the most powerful means of studying specimens at micron and sub micron length scales \cite{Gab01}. Most holographic experiments are performed using beams with a high degree of spatial and temporal coherence to yield greater visibility of the interference images. Naturally, the correlation between coherence and visibility prompts a tendency to devote significant resources to developing new technologies that produce beams with higher coherence. This motivation is compounded, since most quantitative analytical methods rely on the strong assumption that the beam is fully coherent \cite{Pag01}. 

In this study we demonstrate that the partially coherent nature of radiation allows the inference of additional quantitative information of a specimen in the form of depth moments. Under the fully coherent assumption these depth moments cannot be retrieved unless the sample is probed at different orientations. This retrieval of depth moments leads to a form of tomography that does not involve source and sample rotation, either explicitly or implicitly. Moments of scattering distributions provide robust measures of structure, for a wide variety of specimens. For example, Berry and Gibbs \cite{Ber01} showed how to measure mean square specimen densities via inversion of an autocorrelation function from a single micrograph projection, for statistically isotropic scatterers. By considering incident polychromatic plane waves, Fischer and Wolf \cite{Fis01} derived the second moment of the scattering potential for quasi-homogeneous specimens from far-field data, whilst Baliene and Dogariu \cite{Bal01} experimentally obtained the pair-correlation function for such specimens by varying the degree of spatial coherence in a quasi-monochromatic incident beam. In a different context, Teague described how images can be decomposed in terms of moments, a finite number of which suffice for reconstruction of salient image features \cite{Tea02}.  To advance automated pattern recognition, Teague also showed how to efficiently measure transverse moments of images from transparencies using laser interferometry \cite{Tea03}. In this work we describe how to measure longitudinal moments of specimens along the propagation direction of the incident radiation, in a through-focus imaging context.      

Consider an object that is composed of a single material with complex refractive index $n=1-\delta+i\beta$, projected thickness $\textup{T}_{\theta}(x,z=0)$ and illuminated with paraxial quasimonochromatic radiation. Note, only one transverse dimension $x$ will be utilized for simplicity. Then, for sufficiently small object--to--detector distance $z$ which is located downstream from the object, the propagated spectral density $S$ be given by:   

\begin{align} 
S(x,z)=1+(z{k}^{-1}\delta\partial_{x}^{2} -\mu)\left \langle \textup{T}_{\theta} (x,z=0)\right \rangle_{\theta},
\label{Eq2}
\end{align}

\noindent Here, the linear absorption coefficient is $\mu=2k\beta$ where $k$ is the wavenumber. The symbols $\left \langle  \right \rangle_{\theta}$ denote the ensemble average over a range of angles $\theta$, and $\partial^{2}_{x}$ represents the 2$^{\textup{nd}}$ order derivative with respect to $x$. Note that Eq.~\ref{Eq2} is a special case of the more generalized form derived by Beltran $et$ $al.$.~\cite{Bel01} for arbitrary forms of aberrated imaging systems. Equation~\ref{Eq2} can be inverted to retrieve $\left \langle \textup{T}_{\theta} (x,z=0)\right \rangle_{\theta}$ via:     

\begin{align} 
\left \langle \textup{T}_{\theta} (x,z=0)\right \rangle_{\theta}=\mathcal{F}^{-1}\frac{1}{z{k}^{-1}\delta k_{x}^{2} +\mu}\mathcal{F}\left \{ 1-S(x,z) \right \}
\label{Eq3}
\end{align}

\noindent where, $\mathcal{F}$ and $\mathcal{F}^{-1}$ denote forward and inverse Fourier transforms, respectively.

From a physical view point $\left \langle \textup{T}_{\theta} (x,z=0)\right \rangle_{\theta}$ describes the average projected thickness over an ensemble of angular directions, which make an angle $\theta$ with respect to the optic axis $z$. This average can be expressed as:

\begin{align} 
\left \langle \textup{T}_{\theta} (x,z=0)\right \rangle_{\theta}=\int_{-\epsilon}^{\epsilon}h(\theta)\textup{T}_{\theta} (x,z=0)d\theta
\label{Eq4}
\end{align}

\noindent where, $h(\theta)$ represents the probability distribution of the various wavefield directions $\textbf{k}_{\theta}$ of the incoming beam that parametrizes all members of the statistical ensemble. Here, all wavefield directions $\textbf{k}_{\theta}$ make an angle $\theta$ about the $z$--axis.

\begin{figure}[h]
\centering
\includegraphics[scale=0.85]{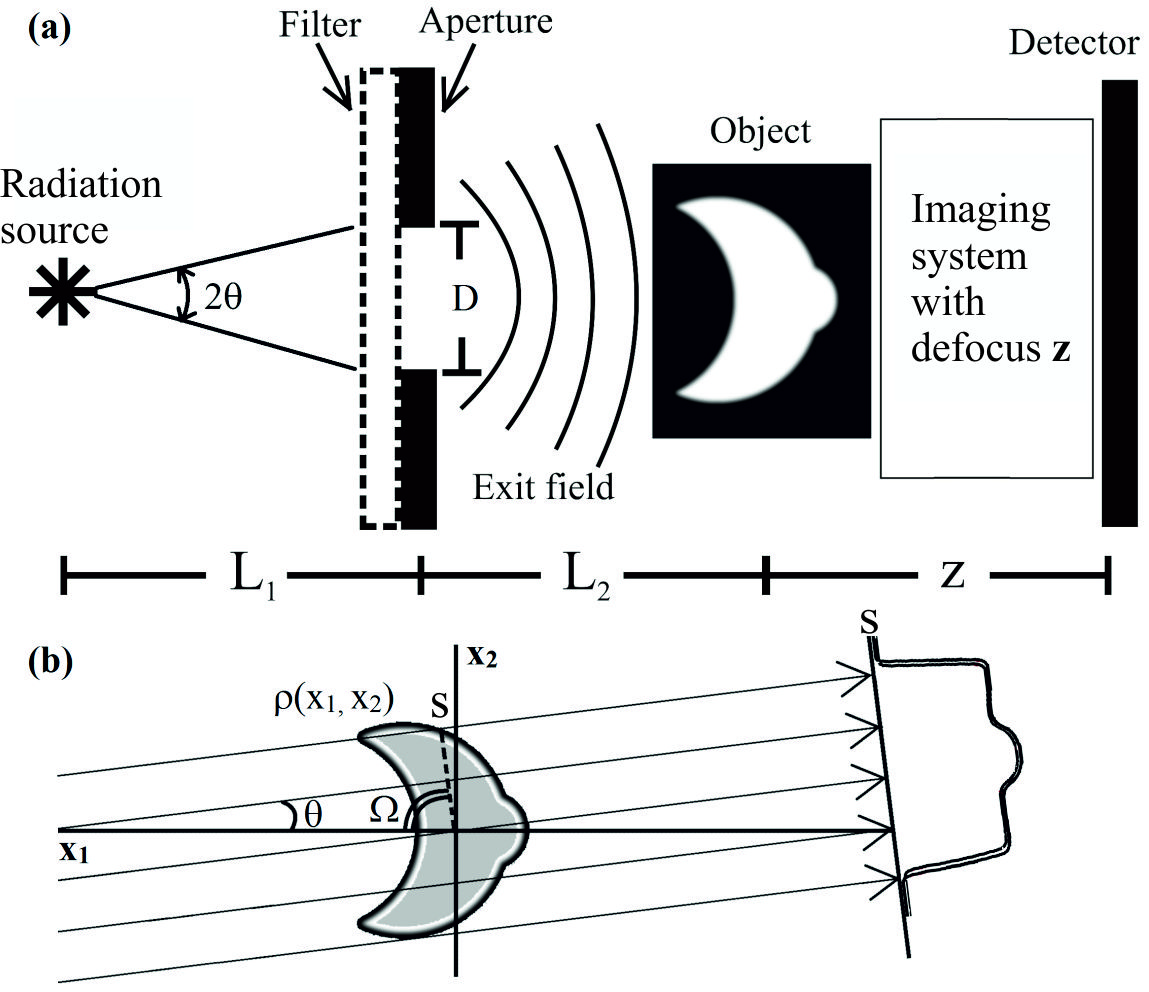}
\caption{a) Imaging setup to extract depth moments from a sample. b) For a particular direction of radiation, parallel line integrals project the object density function $\rho(x_{1},x_{2})$ onto the detector. $s$ and $\Omega$ are polar coordinates. The partially coherent source is composed of an ensemble of radiation directions indexed by $\theta$.}\
\label{figure1}
\end{figure}

The function $\textup{T}_{\theta} (x,z=0)$ can be described by a series of line integrals along the object's density function at many angular orientations (i.e. the Radon transform). Consider the diagram in Fig.~\ref{figure1}b, which portrays a two–-dimensional density function $\rho(x_{1},x_{2})$ with a series of parallel lines used to define integrals along ray paths that comprise a projection of $\rho(x_{1},x_{2})$ at a given angle. The variable $s$ is the shortest distance from the origin to the parallel lines. The equation of each line can be parametrised in terms of the unit normal $\hat{\textup{n}} = (\cos \Omega, \sin \Omega)$ to give the standard form \cite{Kak01}.

\begin{align} 
x_{1} \cos \Omega + x_{2} \sin \Omega = s.
\label{Eq5}
\end{align}

With reference to Fig.~\ref{figure1} (b) we can express Eq.~\ref{Eq5} in terms of the angle between the $x_{1}$ axis and projection $z$--axis with the substitution $\Omega = \pi/2 - \theta$. With this geometry the projected thickness, in the form of a Radon transform, is 

\begin{align} 
\textup{T}_{\theta}(s) = \iint_{-\infty}^{\infty} \rho(x_{1},x_{2}) \hat{\delta }(s-[x_{1} \sin \theta + x_{2} \cos \theta])dx_{1} dx_{2}. 
\label{Eq6}
\end{align}

\noindent Here $ \hat{\delta }$ is the Dirac delta function. Now consider a partially coherent beam with a $h(\theta)$ that comprises a small angular range, that is $\theta \ll 1$, such that Eq.~\ref{Eq5} asymptotically becomes $x_{1}\theta +x_{2} = s$.  Invoking the Dirac delta sifting property on the $x_{2}$ variable, the integrand in Eq.~\ref{Eq6} becomes $\rho(x_{1},s-x_{1}\theta)$ and we are left with an integral over $x_{1}$. In this small $\theta$ limit, the two coordinate systems in Fig.~\ref{figure1} align, to identify $x_{1} = z$ and distances on the detector as $s = x$. Taylor expand about $x$, such that Eq.~\ref{Eq6} becomes,

\begin{align} 
\textup{T}_{\theta}(x)=\sum_{j} \frac{(-\theta)^{j}}{j!}  \partial_{x} ^{j} \int_{-\infty}^{\infty}z^{j} \rho(x,z) dz.
\label{Eq7}
\end{align}

Each member of the light field's statistical ensemble, characterized by the angle of propagation $\theta$, probes a different path through the thickness which is angle--dependent. Moreover, Eq.~\ref{Eq7} defines this angle dependence in the $\left | \theta \right | \ll 1 $ limit as a series of longitudinal moments along the average propagation direction or $z$-axis, with increasing orders of transverse spatial derivatives acting on successively higher orders of moment. These moments represent depth information of the specimen density; the 0$^{\textup{th}}$ moment is the projected thickness at $\theta = 0$, the 1$^{\textup{st}}$ order moment is proportional to the centroid of $\rho(x,z)$ and the 2$^{\textup{nd}}$ moment is related to the moment of inertia etc. In this context, perfectly coherent illumination oriented along the $z$-axis would yield the 0$^{\textup{th}}$ moment and no other information. Hence, whilst partial coherence degrades the visibility of propagation--induced interference fringes, members of the statistical ensemble provide additional information about the specimen as a form of micro--tomography evidenced by the hierarchy of $\theta$--induced moments. Indeed, suppose that all $z$--moments are measured and the transverse derivatives have somehow been inverted. One could then construct a characteristic function in a space Fourier-conjugate to $z$, the inverse Fourier transform of which would reconstruct $\rho(x,z)$ for each $x$ \cite{Abra01}. It remains to assign statistical weights to members of the ensemble and average over the angular distribution,

\begin{align} 
\left \langle \textup{T}_{\theta} (x)\right \rangle_{\theta}&=\sum_{j}h_{\theta}^{j}\mathfrak{L}_{j}\rho_{z}^{j},  \nonumber\\
\textup{where}, h_{\theta}^{j}&=\int_{-\epsilon}^{\epsilon}h(\theta)(-\theta)^{j}d\theta, \nonumber\\
\mathfrak{L}_{j}=\frac{1}{j!} \partial_{x} ^{j},\: \rho_{z}^{j}&=\int_{-\infty}^{\infty}z^{j} \rho(x,z) dz. 
\label{Eq8}
\end{align}

\noindent Recalling Eq.~\ref{Eq2}, Eq.~\ref{Eq8} shows that longitudinal moments of $\rho(x,z)$ can be experimentally measured from propagation induced contrast in the spectral density, and separated if the moments of $h(\theta)$ are known for a sufficiently well characterised source. To this end, consider a variation of the source moments $h_{\theta}^{j}$ created by changing, say, the physical limits of all $\theta$ that contribute to the spectral density to define a set of $n$ angular ranges such that $\theta$ spans $\pm\epsilon_{n}$ for the n$^{\textup{th}}$ range. Upon retrieving $\left \langle \textup{T}_{\theta} (x,z=0)\right \rangle_{\epsilon_{n}}$ via Eq.~\ref{Eq2}, and using a measured set of j$^{\textup{th}}$ order moments for each of the $n$ sources, $h_{\epsilon_{n}}^{j}$, Eq.~\ref{Eq8} can be structured in matrix form:

\begin{align}
\left[ \begin{array}{c} \left \langle \textup{T}_{\theta} (x)\right \rangle_{\epsilon_{1}} \\ \left \langle \textup{T}_{\theta} (x)\right \rangle_{\epsilon_{2}} \\ \vdots  \\ \left \langle \textup{T}_{\theta} (x)\right \rangle_{\epsilon_{n}}  \end{array} \right]= \begin{bmatrix} h_{\epsilon_{1}}^{0} & h_{\epsilon_{1}}^{1} & \cdots  &h_{\epsilon_{1}}^{j} \\ h_{\epsilon_{2}}^{0} & h_{\epsilon_{2}}^{1} & \cdots  & h_{\epsilon_{2}}^{j} \\
\vdots & \vdots  & \ddots& \vdots  \\ h_{\epsilon_{n}}^{0} &h_{\epsilon_{n}}^{1} & \cdots  & h_{\epsilon_{n}}^{j}  \end{bmatrix}  \left[ \begin{array}{c} \mathfrak{L}_{0}\rho_{z}^{0} \\ \mathfrak{L}_{1}\rho_{z}^{1} \\ \vdots  \\ \mathfrak{L}_{j}\rho_{z}^{j} \end{array} \right]
\label{Eq9}
\end{align}

To measure a 2$^{\textup{nd}}$ order longitudinal moment of the density, Eq.~\ref{Eq9} implies that a minimum of three angular ranges for the source distribution is required. However for a symmetric source, all odd moments of $h(\theta)$ are naturally zero, hence only two variations of $\epsilon$ are required in that context. Similarly, whilst no purely odd source distribution is possible since $h(\theta)$ is necessarily positive-definite, a highly asymmetric $h(\theta)$ may comprise a predominant 1$^{\textup{st}}$ order moment $h_{\epsilon_{1}}^{1}$. In that case, Eq.~\ref{Eq9} can also be approximated as a 2 $\times$ 2 matrix equation to retrieve $\mathfrak{L}_{1}\rho_{z}^{1}$, which yields a desired 1$^{\textup{st}}$ order longitudinal moment $\rho_{z}^{1}$ of the specimen density $\rho(x,z)$ after integration along the $x$-axis. 

For a field characterised by two angular degrees of freedom $\theta$ and $\chi$, incident upon a three dimensional (3D) object $\rho(x_{1},x_{2},x_{3})$, we require two versions of Eq.~\ref{Eq5} for two parameters $s_{1}$ and $s_{2}$, which we identify with planar detector coordinates $x$ and $y$. The product of two Dirac delta functions filters $x_{2}$ and $x_{3}$ in a 3D form of Eq.~\ref{Eq6} such that a corresponding Taylor series in $x$ and $y$ for $\left | \theta \right | \ll 1 $, $\left | \chi \right | \ll 1$, yields the following generalisation of Eq.~\ref{Eq8},

\begin{align} 
&\left \langle \textup{T} (x,y)\right \rangle_{\theta,\chi}=\sum_{j,l}h_{\theta,\chi}^{j,l}\mathfrak{L}_{j,l}\rho_{z}^{j,l}, \nonumber\\
\textup{where}, h_{\theta,\chi}^{j,l}&=\int_{-\eta}^{\eta}\int_{-\epsilon}^{\epsilon}h(\theta,\chi)(-\theta)^{j}(-\chi)^{l}d\theta d\chi, \nonumber\\
\mathfrak{L}_{j,l}&=\frac{1}{j!l!} \partial_{x}^{j}\partial_{y}^{l}, \rho_{z}^{j,l}=\int_{-\infty}^{\infty}z^{j+l} \rho(x,y,z) dz. 
\label{Eq10}
\end{align}

For homogeneous objects, with uniform refraction and attenuation properties, the average thickness and 1$^{\textup{st}}$ longitudinal moment can be used to reconstruct the 3D (or 2D) shape, provided that the thickness is well defined (not multivalued along the average propagation direction). Likewise, a broader class of other objects can be completely reconstructed using the 0$^{\textup{th}}$, 1$^{\textup{st}}$ and 2$^{\textup{nd}}$ moments alone; and so on. In the simulations that follow, we have chosen a non-trivial 2D homogeneous object to demonstrate this point.


Refer once again to the imaging setup in Fig.~\ref{figure1}. The filter acts as a monochromator to achieve a partially coherent exit field of mean wavelength $\lambda$. We assume the aperture size $\textup{D}$ is adjustable to allow the coherence of the exit field to be varied. Here, the aberrated imaging system has only a defocus type aberration present, which is the propagation distance $z$ between the object and the detector.       

\begin{figure}[h]
\centering
\includegraphics[scale=0.60]{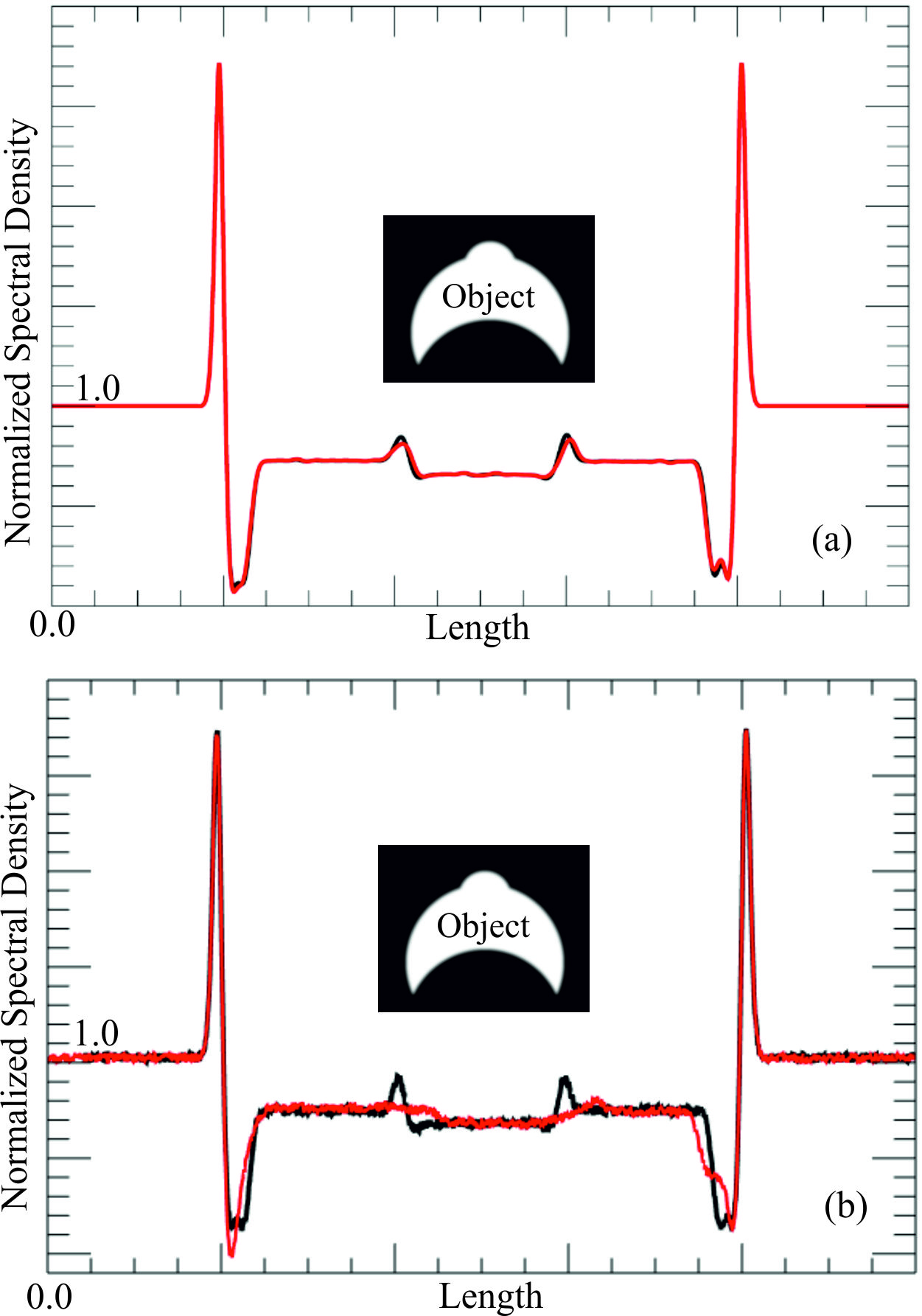}
\caption{Spectral density intensity profiles of forward simulations of a partially coherent field illuminating the object in Fig.~\ref{figure1} (scaled inset) using different angular ranges. (a) The black profile was generated using $\epsilon_{1}=6.0$ $\mu$rad and the red profile was generated using $\epsilon_{2}=9.0$ $\mu$rad. (b) The black profile was generated using $\epsilon_{1}=0.3$ $\mu$rad and the red profile was generated using $\epsilon_{2}=30.0$ $\mu$rad.} 
\label{figure2}
\end{figure}

To simulate the spectral density given an ensemble of monochromatic plane waves, we first separately calculated the exit wavefield $\Psi_{\theta_{m}}(x,z=0)$ for each discrete orientation $\theta_{m}$. The exit wavefield was calculated using the projection approximation with $\delta=4\times 10^{-7}$ and $\beta=1\times 10^{-9}$ with 11 keV photons to satisfy the weak phase-amplitude condition at X-ray wavelengths. The crescent-shaped simulated object was 0.6 mm across in the transverse direction. Each exit wavefield was propagated a defocus distance $z=0.5$ m to obtain $\Psi_{\theta_{m}}(x,z)$, using the angular-spectrum formalism for each discrete orientation $\theta_{m}$ \cite{Pell01}. Upon propagating all wavefields in the ensemble, the cross-spectral density was calculated with

\begin{align} 
\textup{W}(x_{1},x_{2},z)=\sum _{m}h(\theta_{m})\Psi^{*}_{\theta_{m}}(x_{1},z)\Psi_{\theta_{m}}(x_{2},z).
\label{Eq11}
\end{align}

For our simulations we used a probability distribution of the form $h(\theta)=\frac{1}{2\epsilon^{2} }\theta+\frac{1}{2\epsilon}$ to emulate an asymmetric source. The output spectral density was obtained by taking the diagonal of Eq.~\ref{Eq11}; that is, $S(x,z)=\textup{W}(x,x,z)$. Pairs of intensity profiles of this spectral density are shown in Fig.~\ref{figure2} (a) and (b). In Fig.~\ref{figure2} (a) the black and red curves were calculated using the respective angular range of $\epsilon_{1}=6.0$ $\mu$rad and $\epsilon_{2}=9.0$ $\mu$rad. These values were chosen to mimic intermediate to high spatial coherence. Here both curves are almost indistinguishable except in the inner fringes were we can see a slight degradation in visibility. The angular range for the black curve in Fig.~\ref{figure2} (b) was $\epsilon_{1}=0.3$ $\mu$rad to mimic an X-ray beam exhibiting a high degree of spatial coherence. Here, Fresnel fringes are clearly visible at the sharp edges along the transverse direction of the object as a result of free space propagation. In contrast, the red curve was simulated using $\epsilon_{2}=30.0$ $\mu$rad, which is 100$\times$ larger than that used to calculate the red curve. This emulates an X-ray beam with a substantially diminished degree of spatial coherence and can be achieved in practice by increasing the size $\textup{D}$ of the aperture. 0.5$\%$ Poisson noise was added to each spectral density profile in (b).  

\begin{figure}[h]
\centering
\includegraphics[scale=0.50]{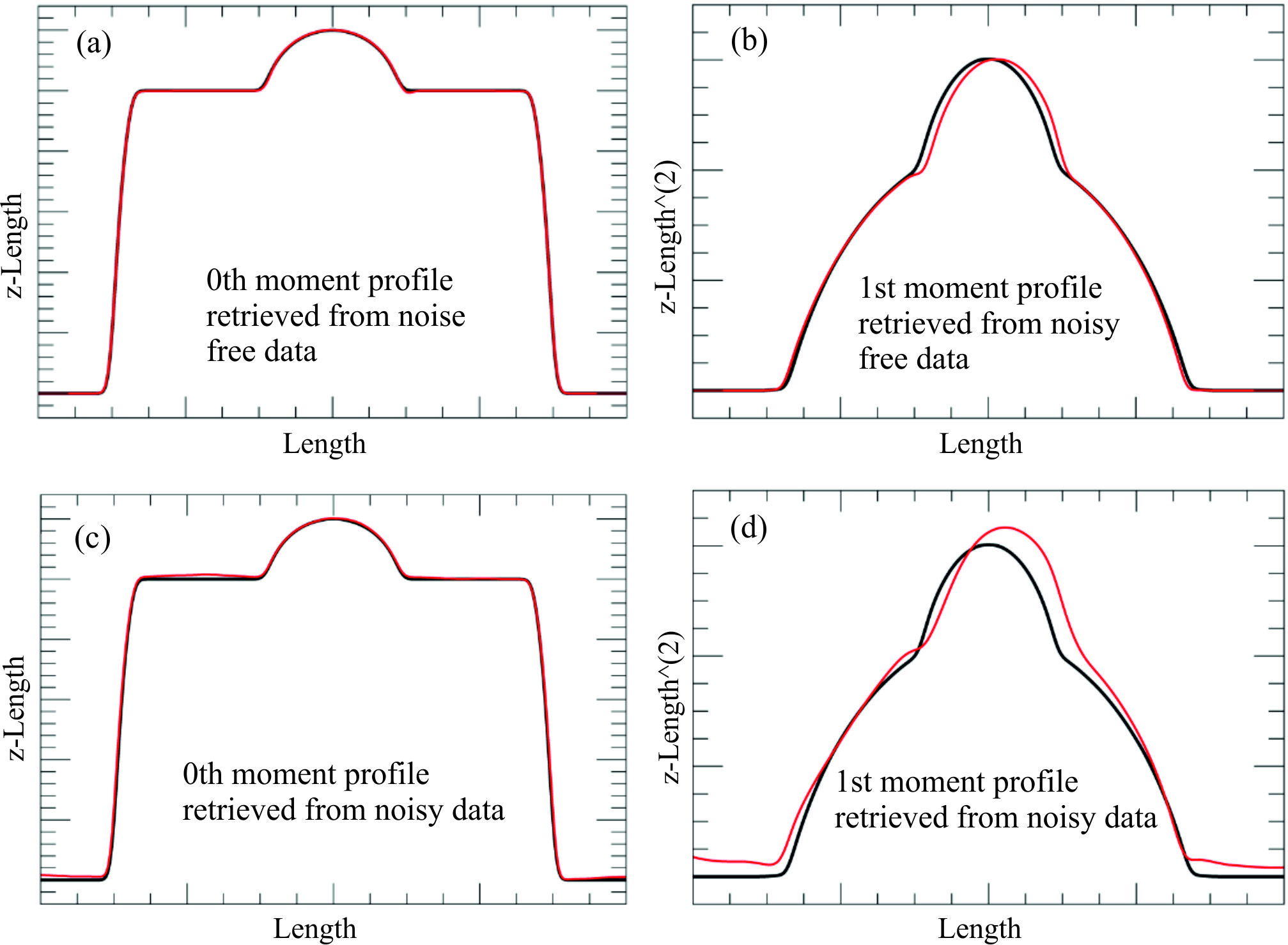}
\caption{(a) and (b) show the recovered zero and first order moments of the object in Fig.~\ref{figure1} using the noise-free spectral densities generated in Fig.~\ref{figure2} (a). The retrieved moments are the red curves, which are overlaid with profiles of the numerically exact moments (black curves). (c) and (d) were recovered using spectral densities generated in Fig.~\ref{figure2} (b) with 0.5$\%$ added Poisson noise.} 
\label{figure3}
\end{figure}

\begin{figure}[h]
\centering
\includegraphics[scale=0.50]{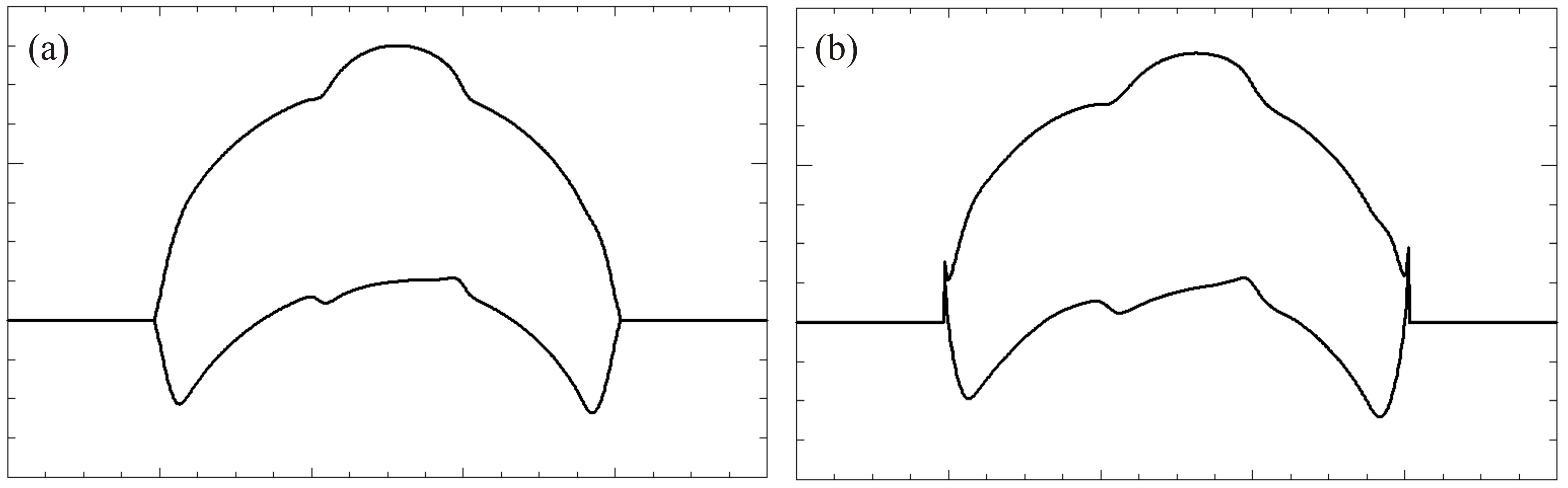}
\caption{Tomographic reconstruction of the object in Fig.~\ref{figure1} using retrieved moments in Fig.~\ref{figure3}. (a) is reconstructed from the moments in Fig.~\ref{figure3} (c) and (d). (b) is reconstructed from the moments in Fig.~\ref{figure3} (a) and (b). Each panel represents a pair of overlaid graphs; one for the top surface and another for the bottom.} 
\label{figure4}
\end{figure}

Since the angular ranges $\epsilon_{n}$ used here are substantially small, moments higher than first order are negligible. Accordingly, Eq.~\ref{Eq9} was solved as a 2$\times$2 matrix equation. Figure~\ref{figure3} shows profiles of the retrieved zero $\rho_{z}^{0}$ and first $\rho_{z}^{1}$ order moments.       
\noindent Lastly, Fig.~\ref{figure4}a and Fig.~\ref{figure4}b display the tomographic reconstruction of the object in the respective absence and presence of noise. These delineations of the specimen boundaries were obtained by respectively adding and subtracting half of the 0$^{\textup{th}}$ moment from the 1$^{\textup{st}}$ moment normalized by the 0$^{\textup{th}}$ moment, to create the upper and lower curves in each panel.  

Before commenting on the simulation findings, it is worth interpreting the broader implications of the theory described here. Considering a 3D specimen density, Eq.~\ref{Eq3} extends to a generalized form of in-line holography \cite{Gab01}, for monochromatic realisations of the partially coherent incident field, representing a solution of an effective transport of intensity equation (TIE) for weakly scattering objects \cite{Teague01}. Further utilizing Eqs.~\ref{Eq9} and \ref{Eq10}, the $\mathfrak{L}_{j,l}\rho_{z}^{j,l}$ can similarly be retrieved and then transversely integrated to yield the desired specimen moments $\rho_{z}^{j,l}$. Note that, for a one-dimensional (1D) source $h(\theta,\chi) \equiv \hat{\delta}(\chi)h(\theta)$, Eq.~\ref{Eq10} reduces to the simpler form of Eq.~\ref{Eq8}, albeit with density $\rho(x,y,z)$. Hence for 3D objects it would appear advantageous to restrict experiments to solely employ 1D sources. However, whilst the linear system in Eq.~\ref{Eq9} is more challenging to solve for $\mathfrak{L}_{j,l}\rho_{z}^{j,l}$ in 3D, a 2$^{\textup{nd}}$ transverse dimension offers increased numerical stability for the subsequent transverse integrations to recover each $\rho_{z}^{j,l}$ \cite{Anis01}.  

In some contexts, the transverse gradients of the moments $\rho_{z}^{j,l}$ may be of direct physical interest and transverse integration of $\mathfrak{L}_{j,l}\rho_{z}^{j,l}$ may then be undesirable. For example, consider coherent fast electrons, which suffer phase changes in proportion to projected electromagnetic potentials for weak fields, as a consequence the Aharonov-Bohm effect \cite{Aha01}. In this context, the coherent defocus induced intensity change is then proportional to the Laplacian of the projected specimen potential. For coherent plane wave electrons, the Laplacian of the 0$^{\textup{th}}$ specimen moment provides the projected charge density, on account of Poisson's equation, which is a well known result in electron microscopy \cite{Lyn01}. For a statistical ensemble of such waves from a symmetric source, defined by a circular angle-limiting aperture, our formalism connects the $z$-propagated spectral density to the transverse Laplacian of the ensemble averaged potentials through a 2D form of Eq. \ref{Eq2}. In turn, $\mathfrak{L}_{2,0}\rho_{z}^{2,0}+\mathfrak{L}_{0,2}\rho_{z}^{0,2}$ represents the 2$^{\textup{nd}}$ longitudinal moment of the charge density, since $\rho_{z}^{0,2}$ = $\rho_{z}^{2,0}\equiv\rho_{z}^{2}$. Likewise, the retrieved 0$^{\textup{th}}$ longitudinal moment $\rho_{z}^{0}$ can be weighted and integrated along x and y to give the transverse moments $\rho_{x}^{2}$ and $\rho_{y}^{2}$, respectively. Hence the complete 2$^{\textup{nd}}$ moment of the charge density $<r^{2}>=\rho_{x}^{2}+\rho_{y}^{2}+\rho_{z}^{2}$ can be measured from as few as two images. Often this moment is connected to the diagmagnetic susceptibility through the mean inner potential for strictly spherically symmetric charge distributions \cite{Sal01}. Here we promote the prospect of mapping the variance $<r^{2}>$ of arbitrary electromagnetic charge distributions for weakly refracting objects. Since the specimen need not be rotated, this form of micro-tomography may prove useful for bright-field aberration-corrected experiments of 2D materials at atomic resolution, for specimens such as doped graphene which contain transversely modulated charge densities \cite{Meyer01}. For different specimens cooled below and above the transition temperature temperature, such experiments could provide vital insights for the theory of superconductivity \cite{His01}.

In the simulation results in Fig.~\ref{figure2}, the reduction in coherence is manifested as a degradation of fringe visibility. For this case it is more evident in the fringes produced at inner edges of the object where they have been almost completely blurred out. It is this variation in fringe blurring which conveys the desired depth moment information. Figure~\ref{figure3} shows that the 0$^{\textup{th}}$ and 1$^{\textup{st}}$ moments were reliably recovered from the data in Fig.~\ref{figure2}, with some deformation of the 1$^{\textup{st}}$ moment due to noise-induced artefacts arising from the Fourier inversions required in Eq.~\ref{Eq3} and Eq.~\ref{Eq8}. Lastly, Fig.~\ref{figure4} shows that the top and bottom surfaces of the object can be faithfully reconstructed. The apparent deformations of the tomogram in Fig.~\ref{figure4}b due to noise might be countered by improved regularization in each Fourier inversion, however an in depth exploration of optimal numerics in this context is outside the scope of this paper.  

In the context of interferometry, such as in-line holography or phase contrast imaging, partial coherence is often deemed to be detrimental, as the resulting loss of fringe visibility appears to degrade interference information. On the contrary, our theoretical study shows that a partially spatially coherent source can provide new information about weakly scattering specimens in the form of moments about the axis of propagation, which conveys robust measures of depth structure. As a byproduct of these insights, the proof-of-principle simulations reported here demonstrated an efficient (two-image) form of in-line holographic tomography, which does not require rotation. This moment retrieval scheme may prove useful for low-dose 3D imaging with partially spatially coherent X-ray sources.

The framework developed here may be compared to, and contrasted with, the formalism based on the TIE \cite{Teague01, Gur01, Gur02, Gur03, Pag02}.  In its typical form, the TIE formalism assumes the projection approximation, and therefore does not facilitate the recovery of depth information without sample rotation.  The effects of partial coherence can be incorporated into a TIE analysis \cite{Pag02, Pet01} but the formalism remains one in which depth information is not obtained, unless the sample is physically rotated.  Broadly speaking, partially-coherent TIE analyses view partial coherence as a factor that must be accounted for so as to properly recover projected quantities, without making use of such partial coherence to extract depth information.  Similar remarks can be made regarding other methods where the source or detector is shaped, such as differential phase contrast \cite{Dek01, Chap01, Anis01, Meh01, Jon01, Mor01, Mor02}.  All of the above may be contrasted with the moment-recovery formalism of the present paper, which exploits the additional information furnished via coherence variation to extract longitudinal depth moments without the need for sample rotation.

As to the practical limitations of our means for retrieving depth moments via coherence variation, the precision with which one can determine the varying coherence properties of the source is an obvious and natural practical limitation.  Moreover, the difference in coherence properties (e.g. source size, angular distribution of ensemble of fields for each stochastic source etc.) between measurements for our method, must be (i) sufficiently large for there to be a statistically significant difference between the data such that the associated difference signal is meaningful; and (ii) sufficiently small that both of the partially-coherent sources are not so lacking in coherence as to wash out meaningful information in the data.  Another obvious limitation, which could be said to apply to all imaging schemes, is the need for a sufficiently high signal-to-noise ratio in the raw intensity data.  Lastly, we mention that the limitations in the stability of the method can be readily analysed by studying the condition number (ratio of maximum to minimum singular values) of the matrix appearing in Eq. 8.  The closer this condition number is to unity, the more stable the recovery.  This last-mentioned fact will be useful in the future practical application of our method.

\end{document}